\documentclass[onecolumn,preprintnumbers,eqsecnum,superscriptaddress,12pt]{revtex4}
\usepackage{amsfonts}
\usepackage{amssymb}
\usepackage{amsmath}
\usepackage{epsf}
\usepackage{graphicx}
\usepackage{dcolumn}
\usepackage{bm}
\usepackage{color}
\usepackage[colorlinks=true,
            citecolor=blue,
            urlcolor=blue,
            filecolor=black,
            linktocpage=true,
            linkcolor=blue,,
             ]{hyperref}

\newcommand{\beq}{\begin{equation}}
\newcommand{\eeq}{\end{equation}}
\newcommand{\beqs}{\begin{eqnarray}}
\newcommand{\eeqs}{\end{eqnarray}}

\newcommand{\be}{\begin{equation}}
\newcommand{\ee}{\end{equation}}
\newcommand{\bea}{\begin{eqnarray}}
\newcommand{\eea}{\end{eqnarray}}

\def\[{\left[}
\def\]{\right]}
\def\({\left(}
\def\){\right)}

\begin{document}

\title{Gravity/Fluid Correspondence and Its Application on Bulk Gravity with $U(1)$ Gauge Field}
\author{Ya-Peng Hu}\email{huyp@pku.edu.cn}
\address{INPAC, Department of Physics, and Shanghai Key Laboratory of Particle Physics and Cosmology, Shanghai Jiao Tong University, Shanghai 200240, China }
\address{College of Science, Nanjing University of Aeronautics and Astronautics, Nanjing 210016, China}

\author{Jian-Hui Zhang}\email{zhangjianhui@gmail.com}
\address{INPAC, Department of Physics, and Shanghai Key Laboratory of Particle Physics and Cosmology, Shanghai Jiao Tong University, Shanghai 200240, China }

\begin{abstract}
As the long wavelength limit of the AdS/CFT correspondence, the
gravity/fluid correspondence has been shown to be a useful tool for
extracting properties of the fluid on the boundary dual to the gravity in the bulk. In this
paper, after briefly reviewing the algorithm of gravity/fluid
correspondence, we discuss the results of its application on bulk
gravity with a $U(1)$ gauge field. In the presence of a $U(1)$ gauge
field, the dual fluid possesses more interesting properties such as
its charge current. Furthermore, an external field $A_\mu^{ext}$
could affect the charge current, and the $U(1)$ Chern-Simons term
also reinduces extra structures to the dual current giving anomalous
transport coefficients.


\end{abstract}

\maketitle

\vspace*{1.cm}

\newpage

\section{Introduction}
The AdS/CFT
correspondence~\cite{Maldacena:1997re,Gubser:1998bc,Witten:1998qj,Aharony:1999ti}
provides a remarkable connection between a gravitational theory and
a quantum field theory. According to this correspondence, the
gravitational theory in an asymptotically AdS spacetime can be
reformulated in terms of a quantum field theory on its boundary. In
particular, the dynamics of a classical gravitational theory in the
bulk can be mapped into a strongly coupled quantum field theory on
the boundary. Therefore, the AdS/CFT correspondence provides a
useful tool of investigating the strongly coupled field theory from
the dual classical gravitational
theory~\cite{Herzog:2009xv,CasalderreySolana:2011us}.

Since the discovery of the AdS/CFT correspondence, there has been
much work studying the hydrodynamical behavior of the dual quantum
field theory using this
correspondence~\cite{Policastro:2001yc,Kovtun:2003wp,Buchel:2003tz,Kovtun:2004de}.
A particular example is the computation of shear
viscosity\cite{Mas:2006dy,Son:2006em,Saremi:2006ep,Maeda:2006by,Cai:2008in,Brustein:2008cg,Iqbal:2008by,Cai:2008ph,Cai:2009zv,Astefanesei:2010dk,Buchel:2008ae,Myers:2008yi,
Policastro:2002se,Policastro:2002tn,Buchel:2004qq,Cohen:2007qr,Son:2007vk,Cherman:2007fj,Chen:2007jq,Fouxon:2008pz,Dobado:2008ri,Landsteiner:2007bd,
Brigante:2007nu,Brigante:2008gz,Kats:2007mq,deBoer:2009pn,Camanho:2009vw,Buchel:2009sk,Adams:2008zk,Cremonini:2009sy,Ge:2008ni,deBoer:2009gx,Camanho:2009hu}.
The reason that the correspondence applies here is because
hydrodynamics is an effective description of interacting quantum
field theory in the long wavelength limit, i.e. when the length
scales under consideration are much larger than the correlation
length of the quantum field theory. Later, the long wavelength limit
of the AdS/CFT correspondence has been developed as the
gravity/fluid
correspondence~\cite{Bhattacharyya:2008jc,Rangamani:2009xk}, which
could provide a systematic mapping of bulk gravity and boundary
fluid. Through the gravity/fluid correspondence, one can construct
the stress-energy tensor of the fluid order by order from the
derivative expansion of the gravity solution, and then extract
certain transport coefficients of the fluid. For example, the shear
viscosity $\eta$, entropy density $s$, and thus their ratio $\eta/s$
can be calculated from the first-order stress-energy
tensor~\cite{Hur:2008tq, Erdmenger:2008rm, Banerjee:2008th,
Son:2009tf, Tan:2009yg, Torabian:2009qk,
Hu:2010sn,Hu:2011ze,Hu:2011qa,Bai:2012ci}. In addition, in the
presence of extra gauge fields in the bulk, the correspondence also
allows us to construct the conserved charge current in the dual fluid,
and thereby allows us to extract the thermal and electrical
conductivities of the dual fluid~\cite{Hur:2008tq, Erdmenger:2008rm,
Banerjee:2008th}. It should be pointed out that the presence of an
external field $A_{\mu}^{ext}$ is needed if one wants to obtain the
electrical conductivity~\cite{Hur:2008tq,Son:2009tf,
Hu:2011ze,Hu:2011qa}. Furthermore, it has also been shown that
introducing the topological Chern-Simons term in the bulk gravity
brings interesting features such as {the chiral magnetic effect
(CME)} and the chiral vortical effect (CVE) into the dual
fluid~\cite{Son:2009tf, Tan:2009yg, Hu:2011ze,Hu:2011qa,
Kalaydzhyan:2011vx, Amado:2011zx}. The Chern-Simons term was first
discussed in Maxwell theory in three dimensions where it renders the
gauge theory massive~\cite{Deser:1982vy}. Such terms also affect the
phase transition of holographic superconductors in four
dimensions~\cite{Tallarita:2010uh} and stability of the
Reissner-Nordstrom (RN) black holes in AdS space in five dimensions
\cite{Nakamura:2009tf}. In this paper, we will discuss the results
for transport coefficients of dual boundary fluid in the presence of
a $U(1)$ gauge field in the bulk.

The rest of this paper is organized as follows. We start Sec.II by
briefly reviewing the algorithm of gravity/fluid correspondence. In Sec.~III, we
discuss the application of gravity/fluid correspondence to
studying fluid dynamics dual to bulk gravity in the presence of a
$U(1)$ gauge field. One can see that both the stress tensor and the conserved charge current of the dual fluid can be extracted. Moreover, adding the external field $A_{\mu}^{ext}$ and
Chern-Simons term in the bulk also introduces extra structures to the charge current of
dual fluid. The conclusion and discussion are given in Sec.~IV.

\section{Algorithm of gravity/fluid correspondence}
After several years' development, there exists now a well-formulated
algorithm of how to extract information on the boundary fluid from
the dual gravity using the gravity/fluid correspondence. In this
section, we just give a brief introduction, and the more details
could be seen in~\cite{Bhattacharyya:2008jc,Rangamani:2009xk}. For
illustrative purposes, let us take the $5$-dimensional Einstein
gravity in the bulk as an example. Its action can be written as
\begin{eqnarray}
\label{action} I=\frac{1}{16 \pi G}\int_\mathcal{M}~d^5x
\sqrt{-g^{(5)}} \left(R-2 \Lambda \right)\ , \label{ExAction}
\end{eqnarray}
where $R$ is the Ricci scalar, and $\Lambda$ is a negative
cosmological constant $\Lambda=-6/\ell^2$. For later convenience, we
set $\ell=1$ and $16\pi G=1$. The equation of motion then reads
\begin{eqnarray}
\label{eqs} R_{MN } -\frac{1}{2}Rg_{MN}+\Lambda
g_{MN}&=&0~.\label{ExEqu}
\end{eqnarray}
The starting point is the black brane solution written in the
Eddington-Finkelstein coordinate
\begin{equation}
ds^2=g_{MN}dx^Mdx^N=2H(r,b)dv dr-r^2 f(r,b)dv^2+r^2(dx^2+dy^2+dz^2)\ ,
\label{ExBlackbrane}
\end{equation}
where $x^M=(v,x_{i},r)$, $b$ is a constant related to the mass of
the black brane. This black brane solution is asymptotical to the
$AdS_5$ solution, which means $H(r)\rightarrow 1, f(r)\rightarrow 1$
when $r$ approaches infinity. The advantage of using the
Eddington-Finkelstein coordinate system is that it avoids the
coordinate singularity.

By boosting the black brane solution in Eq.~(\ref{ExBlackbrane}),
one obtains a solution with more parameters which can then be
related to the degrees of freedom of the dual boundary fluid
\begin{equation}\label{EXBBboost}
ds^2 =2H(r,b) u_\mu dx^\mu dr - r^2 f(r,b)( u_\mu dx^\mu )^2 +
r^2P_{\mu \nu} dx^\mu dx^\nu
\end{equation}
with
\begin{equation}
u^v = \frac{1}{ \sqrt{1 - \beta_i^2} }~~,~~u^i = \frac{\beta_i}{ \sqrt{1 - \beta_i^2} },~~P_{\mu \nu}= \eta_{\mu\nu} + u_\mu u_\nu\ ,
\end{equation}
where $x^\mu=(v,x_{i})$, velocities $\beta^i $ are constants,
$P_{\mu \nu}$ is the projector onto spatial directions, and the
indices in the boundary are raised and lowered with the Minkowsik
metric $\eta_{\mu\nu}$. The metric~(\ref{EXBBboost}) describes the
uniform boosted black brane moving at velocity $\beta^i
$~\cite{Bhattacharyya:2008jc}.

The transport coefficients of the dual fluid can be extracted by
perturbing the system away from equilibrium. On the gravity side,
this can be achieved by promoting the parameters in the boosted
black brane solution (\ref{EXBBboost}) to functions of boundary
coordinates $x^\mu$. Since the parameters now depend on the boundary
coordinates, the metric~(\ref{EXBBboost}) is no longer a solution of
the equation of motion~(\ref{ExEqu}), extra correction terms are
needed to make the new metric a solution. It is useful to define the
following tensor
\begin{eqnarray}
\label{EXTensors} &&W_{IJ} = R_{IJ} + 4g_{IJ}\ ,
\end{eqnarray}
which arises from the left hand side of Eq.~(\ref{ExEqu}). When the
parameters become functions of boundary coordinates $x^\mu$, $W_{\mu
\nu}$ no longer vanishes and is proportional to the derivatives of
the parameters. These terms are the source terms which will be
canceled by extra correction terms introduced into the metric. If we
expand the parameters around $x^\mu=0$ to the first order
\begin{eqnarray}
\beta_i&=&\partial_{\mu} \beta_{i}|_{x^\mu=0}
x^{\mu},~~~b=b(0)+\partial_{\mu} b|_{x^\mu=0} x^{\mu}\ ,
\label{EXExpand}
\end{eqnarray}
where we have assumed $\beta^i(0)=0$, we find the first-order source terms by
inserting the metric (\ref{EXBBboost})
with~(\ref{EXExpand}) into $W_{IJ }$. By choosing an appropriate gauge like the background field gauge in \cite{Bhattacharyya:2008jc} ($G$
represents the metric)
\begin{equation}
G_{rr}=0,~~G_{r\mu}\propto
u_{\mu},~~Tr((G^{(0)})^{-1}G^{(1)})=0\ ,\label{gauge}
\end{equation}
and taking into account the spatial $SO(3)$ symmetry preserved in
the background metric~(\ref{ExBlackbrane}), the first-order
correction terms around $x^\mu=0$ can be written as
\begin{eqnarray}\label{EXcorrection}
&&{ds^{(1)}}^2 = \frac{ k(r)}{r^2}dv^2 + 2 h(r)dv dr + 2
\frac{j_i(r)}{r^2}dv dx^i +r^2 \left(\alpha_{ij}(r) -\frac{2}{3}
h(r)\delta_{ij}\right)dx^i dx^j\ .
\end{eqnarray}
The first order perturbative solution can then be obtained by
requiring a cancelation of the source terms and the correction
terms. Note that, after obtaining the solution around $x^\mu=0$, one
could convert it into a covariant form so that it applies to other
spacetime points \cite{Bhattacharyya:2008jc}.

Given the first order perturbative solution for the bulk gravity, we are able to extract information of the dual fluid using the gravity/fluid correspondence. According to the correspondence, the stress tensor of dual fluid $\tau
_{\mu\nu}$ can be obtained from the following
relation~\cite{Myers:1999psa}
\begin{eqnarray}\label{EXrelation}
\sqrt{-h}h^{\mu\nu}<\tau _{\nu\sigma}>=\lim_{r\rightarrow \infty
}\sqrt{-\gamma }\gamma ^{\mu\nu}T_{\nu\sigma}.
\end{eqnarray}
where $h^{\mu\nu}$ is the background metric upon which the dual field theory resides, $\gamma ^{\mu\nu}$ is the boundary metric obtained from
the well-known ADM decomposition
\begin{eqnarray}
ds^2 = \gamma_{\mu\nu}(dx^\mu + V^\mu dr)(dx^\nu + V^\nu dr) + N^2 dr^2\ ,
\end{eqnarray}
and $T_{\mu\nu}$ is the boundary stress tensor defined as
\begin{equation}
T_{\mu\nu}=\frac{2}{\sqrt{-\gamma }}\frac{\delta }{\delta \gamma
^{\mu\nu}}\left( I+I_{\mathrm{sur}}+I_{\text{ct}}^0 \right),
\label{EXTab}
\end{equation}
where
\begin{equation}
I_{\mathrm{sur}}=-\frac{1}{8\pi G}\int_{\partial
\mathcal{M}}d^{4}x\sqrt{-\gamma }K \label{EXSurfaceterm}
\end{equation}
is the Gibbons-Hawking surface term, $K$ is the trace of the
extrinsic curvature $K_{\mu\nu}$ of the boundary, which is given by
$K_{\mu\nu}=-\frac{1}{2}(\nabla_{\mu}n_{\nu}+\nabla_{\nu}n_{\mu})$
with $n^{\mu}$ the normal vector of the constant hypersurface
$r=r_c$ pointing toward increasing $r$ direction. In addition,
\begin{eqnarray}
\label{Lagrangianct} I_{\mathrm{ct}}^0 &=&\frac{1}{8\pi
G}\int_{\partial \mathcal{M}} d^{4}x\sqrt{-\gamma } \left[
  -3
 -\frac{\mathsf{R}}{4}\right]
\end{eqnarray}
is the boundary counterterm, and $\mathsf{R}$ is the curvature
scalar associated with the induced metric on the boundary
$\gamma_{ab}$~\cite{Balasubramanian:1999re,de
Haro:2000xn,Emparan:1999pm,Mann:1999pc}.

From Eq.~(\ref{EXTab}), the boundary stress tensor is
\begin{equation}
 T_{\mu\nu}=\frac{1}{8 \pi G}[K_{\mu\nu}-\gamma _{\mu\nu}K
-3 \gamma _{\mu\nu} +\frac{1}{2}%
( \mathsf{R}_{\mu\nu}-\frac{1}{2}\gamma _{\mu\nu}\mathsf{R})]\ ,
\label{EXTabCFT}
\end{equation}
which can be used to compute the boundary stress tensor of the first
order perturbative black brane solution.

The background metric upon which the dual field theory resides
usually is chosen as $h_{\mu\nu}=\lim_{r \rightarrow \infty}
\frac{\ell^2}{r^2}\gamma_{\mu\nu}$~\cite{Myers:1999psa}, and
\begin{equation}
ds^2=h_{\mu\nu}dx^\mu dx^\nu=-dv^2+dx^2+dy^2+dz^2\ 
\end{equation}
is the Minkowski metric. From
Eq.~(\ref{EXrelation}) and the result for the boundary stress tensor
$T_{\mu \nu}$, the first order stress tensor of the dual fluid $\tau
_{\mu \nu}$ can be found to be
\begin{equation}
\tau_{\mu \nu}=P(\eta_{\mu\nu} + 4 u_\mu u_\nu  ) - 2 \eta
\sigma_{\mu\nu}\ . \label{EXStressTensor}
\end{equation}
from which one immediately reads off the pressure and viscosity of the dual fluid.

\section{Applications on the bulk gravity with $U(1)$ gauge field}
In the previous section, we illustrate the algorithm of the
gravity/fluid correspondence with a simple pure gravity model, which
yields a viscosity for the dual fluid on the boundary. In general,
the gravity/fluid correspondence could be applied to studying fluid
dynamics dual to various bulk gravity configurations with matter
fields in the bulk gravity, and it has been found that after adding
the matter fields to the bulk gravity, more interesting properties
of the dual fluid could be extracted. We discuss here the impact on
the dual boundary fluid of adding a $U(1)$ gauge field to the bulk.

\subsection{The simplest case}
The action with a $U(1)$ gauge field can be written as
\cite{Erdmenger:2008rm,Banerjee:2008th}
\begin{eqnarray}
I &=&\frac{1}{16 \pi G}\int_\mathcal{M}~d^5x \sqrt{-g^{(5)}}
\left(R-2 \Lambda \right)-\frac{1}{4g^2}\int_\mathcal{M}~d^5x
\sqrt{-g^{(5)}}F^2\ ,\label{action1}
\end{eqnarray}
which gives the following equations of motion
\begin{eqnarray}
R_{AB } -\frac{1}{2}Rg_{AB}+\Lambda g_{AB}-\frac{1}{2g^2}\left(F_{A C}{F_{B }}^{C}-\frac{1}{4}g_{AB}F^2\right)&=&0~, \nonumber\\
\nabla_{B} {F^{B}}_{A} &=&0\ . \label{IVeqs1}
\end{eqnarray}
Following the general algorithm described in Sec. II, we start with the five-dimensional charged RN-AdS black
brane solution \cite{Cai:1998vy,Cvetic:2001bk,Anninos:2008sj}
\begin{eqnarray}
ds^2=\frac{dr^2}{r^2f(r)}+r^2
 \left(\mathop\sum_{i=1}^{3}dx_i^2 \right)-r^2f(r) dt^2, \label{IVSolution}
\end{eqnarray}%
where
\begin{eqnarray}
\label{IVf-BH} f(r) &=& 1-\frac{2M}{r^{4}}+\frac{Q^2}{r^6},~~F =
-g\frac{2\sqrt 3 Q}{r^3}dt \wedge dr\ . ~~
\end{eqnarray}%
The outer horizon of the black brane is located at $r=r_{+}$,
where $r_{+}$ is the largest root of $f(r)=0$, and its Hawking
temperature is
\begin{eqnarray}
T_{+}&=&\frac{(r^2f(r))'}{4 \pi}|_{r=r_{+}}=\frac{1}{2 \pi
r_{+}^3}(4M-\frac{3Q^2}{r_{+}^2})\ .\label{IVTemperature}
\end{eqnarray}
Writting the above black brane solution in the Eddington-Finkelstin
coordinate system, one has
\begin{eqnarray}\label{IVSolution1}
ds^2 &=& - r^2 f(r)dv^2 + 2 dv dr + r^2(dx^2 +dy^2 +dz^2)\ ,\\
F &=& -g\frac{2\sqrt 3 Q}{r^3}dv \wedge dr\ , \notag~~
\end{eqnarray}
where $v=t+r_*$, and $r_*$ is the tortoise coordinate satisfying
$dr_*=dr/(r^2f)$. The boosted solution can then be written as
\begin{eqnarray}
ds^2 &=& - r^2 f(r)( u_\mu dx^\mu )^2 - 2 u_\mu dx^\mu dr + r^2 P_{\mu \nu} dx^\mu dx^\nu, \label{IIrnboost}\\
F &=& -g\frac{2\sqrt 3 Q}{r^3} u_\mu dx^\mu \wedge
dr,~~~A=-\frac{\sqrt 3 g Q}{r^2}u_{\mu}dx^{\mu} \label{vector}~~
\end{eqnarray}
with
\begin{equation}
u^v = \frac{1}{ \sqrt{1 - \beta_i^2} }~~,~~u^i = \frac{\beta_i}{
\sqrt{1 - \beta_i^2} },~~P_{\mu \nu}= \eta_{\mu\nu} + u_\mu u_\nu~.
\end{equation}
where velocities $\beta^i $, $M$, $Q$ are constants.

The procedure of solving for the first-order perturbative solution
is essentially the same as in the previous section. Now the tensors
inducing the source terms become
\begin{eqnarray}
&&W_{IJ} = R_{IJ} + 4g_{IJ}+\frac{1}{2g^2}\left(F_{IK}{F^{K}}_J +\frac{1}{6}g_{IJ}F^2\right),\label{IVTensors1}\\
&&W_A = \nabla_{B} {F^{B}}_{A}~.\label{IVTensors2}
\end{eqnarray}
Letting the parameters be $x^\mu$-dependent and expanding them
around $x^\mu=0$ to the first-order, one has
\begin{eqnarray}
\beta_i&=&\partial_{\mu} \beta_{i}|_{x^\mu=0}
x^{\mu},~~~M=M(0)+\partial_{\mu} M|_{x^\mu=0}
x^{\mu},~~~Q=Q(0)+\partial_{\mu} Q|_{x^\mu=0} x^{\mu}\ .
\label{IIExpand}
\end{eqnarray}
and the corresponding required first order correction terms around
$x^\mu=0$ are then given by
\begin{eqnarray}\label{correction}
&&{ds^{(1)}}^2 = \frac{ k(r)}{r^2}dv^2 + 2 h(r)dv dr + 2
\frac{j_i(r)}{r^2}dv dx^i
 +r^2 \left(\alpha_{ij}(r) -\frac{2}{3} h(r)\delta_{ij}\right)dx^i dx^j, \notag\\
&&A^{(1)} = a_v (r) dv + a_i (r)dx^i~.\label{correction1}
\end{eqnarray}
Note that the gauge $a_r(r)=0$ has been chosen. Therefore, the
first-order perturbative gravitational and Maxwell solution could be
achieved by requiring a cancelation of the source terms and the
correction terms.

The solutions then allow one to compute the boundary stress tensor. According to the dictionary of fluid-gravity correspondence, the first-order stress tensor of the dual fluid $\tau _{\mu \nu}$ then reads
\begin{equation}
\tau_{\mu \nu}=\frac{1}{16\pi G}[\frac{2M}{\ell^3}(\eta_{\mu\nu} + 4
u_\mu u_\nu)-\frac{2r_{+}^3}{
\ell^3}\sigma_{\mu\nu}]=P(\eta_{\mu\nu} + 4 u_\mu u_\nu  ) - 2 \eta
\sigma_{\mu\nu} \label{IIIStressTensor}
\end{equation}
with the pressure and viscosity given by
\begin{equation}
P=\frac{M}{8 \pi G \ell^3},~~~\eta=\frac{r_{+}^3}{16 \pi G \ell^3}\ .
\label{IIIets}
\end{equation}
where the coupling $16 \pi G$ and $\ell$ are recovered for
comparison purposes.

Following the AdS/CFT correspondence, the dual operator of
$\tilde A_\mu$ (here $\tilde A_\mu$ is the boundary value of $A_\mu$
projected on the boundary) is the charge current of the dual fluid. One can thus extract the charge current of dual fluid as
\begin{equation}\label{IIIcurrent}
J^\mu = \lim_{r \rightarrow \infty}  \frac{r^4}{\sqrt{-\gamma}}
\frac{\delta S_{cl}}{\delta \tilde A_\mu} = \lim_{r \rightarrow
\infty}  \frac{r^4 N}{g^2}F^{r \mu}~,
\end{equation}
where the factor $r^4$ comes from the conformal transformation.
From this equation and the result for the first-order perturbative solution, one obtains the charge current
\begin{eqnarray}
J^\mu &=& J_{(0)}^\mu + J_{(1)}^\mu\ , \label{IIIfirst order current
Component}
\end{eqnarray}
with the zeroth-order particle number current
\begin{equation}
J_{(0)}^{\mu} = \frac{2\sqrt 3 Q}{g} u^\mu :=nu^\mu\ \label{zero
current},
\end{equation}
and the first-order charge current
\begin{eqnarray}
J_{(1)}^{\mu }=-\kappa P^{\mu \nu }\partial _{\nu}(\frac{\mu }{T}%
)\ . \label{IIIfirst order current}
\end{eqnarray}
where $\mu=\frac{\sqrt 3 gQ}{  r_+ ^2 }$ is the chemical potential,
and $\kappa = \frac{\pi ^2 T^3 r_+^7}{4 g^2M^2}$ is the thermal
conductivity. From this simplest case, it is obvious that after
adding the $U(1)$ gauge field in the bulk, the charge current of
dual fluid could also be extracted.

\subsection{In the presence of an external field $A_{\mu}^{ext}$}
It has been found that in the presence of an external field
$A_{\mu}^{ext}$, the charge current of dual fluid has extra
structures, whereas this external field does not change the stress
tensor of dual fluid~\cite{Hur:2008tq,Son:2009tf,
Hu:2011ze,Hu:2011qa}. The key point is that a constant external
field $A_{\mu}^{ext}$ could be added into (\ref{vector})
like~\cite{Hur:2008tq}
\begin{eqnarray}
A=( A_{\mu}^{ext}-\frac{\sqrt 3 g Q}{r^2}u_{\mu})dx^{\mu}.
\label{Externalfield}~~
\end{eqnarray}
It is obvious Eq.~(\ref{Externalfield}) is still the solution since
the equations of motion in the previous subsection involve the field
strength only.

After requiring the external field $A_{\mu}^{ext}$ to be a function
of $x_{\mu}$, we can expand it around $x_{\mu} = 0$ to the first order as in (\ref{IIExpand}),
\begin{eqnarray}
A_{\mu}^{ext}&=&A_{\mu}^{ext}(0)+\partial_{\nu}
A_{\mu}^{ext}|_{x^\mu=0} x^{\nu}. \label{IIExpandA}
\end{eqnarray}
The corresponding correction terms around $x^\mu=0$ could be same as
(\ref{correction1}). One finds that the external field
$A_{\mu}^{ext}$ changes the coefficients $a_i(r),j_i(r)$ in
Eq.~(\ref{correction1}), and hence change the charge current of dual
fluid~\cite{Hur:2008tq,Hu:2011ze,Hu:2011qa}. Finally, the result of
the first-order charge current now becomes
\begin{eqnarray}\label{IIfamiliar current}
J_{(1)}^\mu = -\kappa  P^{\mu  \nu } \partial _{\nu }\frac{\mu
}{T}+\sigma_{E} u^{\lambda } F ^{\rm ext}{}_{\lambda }{}^{\mu }\ ,
\end{eqnarray}
where
\begin{eqnarray} \label{IITEconductivity} \kappa =  \frac{\pi
^2 T^3 r_+^7}{4 g^2M^2 }\ , \;\;\quad \sigma_{E} =\frac{\pi ^2 T^2
r_+^7}{4g^2 M^2}
 \end{eqnarray}
are the thermal and electrical conductivity,
respectively. Obviously, a new structure related to the electrical
conductivity appears in the first order charge current.

\subsection{Adding $U(1)$ Chern-Simons term in the bulk}
Recently Chern-Simons terms have attracted much attention, as
introducing such terms in the bulk leads to anomalous transport
coefficients on the dual fluid side. The bulk gravity action with a
$U(1)$ Chern-Simons term is
\cite{Erdmenger:2008rm,Banerjee:2008th,Son:2009tf,Hu:2011ze,Hu:2011qa}
\begin{eqnarray}
I &=&\frac{1}{16 \pi G}\int_\mathcal{M}~d^5x \sqrt{-g^{(5)}}
\left(R-2 \Lambda \right)\notag\\
&-&\frac{1}{4g^2}\int_\mathcal{M}~d^5x \sqrt{-g^{(5)}}(F^2
+\frac{4\kappa_{cs} }{3}\epsilon ^{LABCD}A_{L}F_{AB}F_{CD})\
,\label{action1}
\end{eqnarray}
and the equations of motion become
\begin{eqnarray}
R_{AB } -\frac{1}{2}Rg_{AB}+\Lambda g_{AB}-\frac{1}{2g^2}\left(F_{A C}{F_{B }}^{C}-\frac{1}{4}g_{AB}F^2\right)&=&0~,\nonumber \\
\nabla_{B} {F^{B}}_{A}-\kappa_{cs} \epsilon_{ABCDE}F^{BC}F^{DE}
&=&0\ .\label{IVeqs2}~~
\end{eqnarray}
Although the Maxwell equation is different from Eq.~(\ref{IVeqs1}), it turns out that the five-dimensional charged RN-AdS
black brane solution (\ref{IIrnboost})(\ref{vector}) still solves the
above equations (\ref{IVeqs2}). In the following we also insert an external field $A_{\mu}^{ext}$ into the bulk solution. Now the tensors inducing the source terms become
\begin{eqnarray}
&&W_{IJ} = R_{IJ} + 4g_{IJ}+\frac{1}{2g^2}\left(F_{IK}{F^{K}}_J +\frac{1}{6}g_{IJ}F^2\right),\label{IVTensors1}\\
&&W_A = \nabla_{B} {F^{B}}_{A}-\kappa_{cs}
\epsilon_{ABCDE}F^{BC}F^{DE}~.\label{IVTensors2}
\end{eqnarray}
One can solve the first-order perturbative gravitational and Maxwell
solution as before. Note that the Chern-Simons term appears in the
second term of $W_A$ in (\ref{IVTensors2}), while it does not change
$W_{IJ}$ in (\ref{IVTensors1}). As a consequence,
$h(r),k(r),\alpha_{ij}$ are the same as in the case without the
Chern-Simons term, while $a_{i}(r)$ and $j_{i}(r)$ are
different~\cite{Hu:2011ze,Hu:2011qa}. Therefore, the Chern-Simons
term will not change the first-order stress tensor of the dual fluid
$\tau _{\mu \nu}$. However, in the presence of Chern-Simons term,
the charge current of dual fluid can be computed as
\begin{equation}\label{IIIcurrent}
J^\mu = \lim_{r \rightarrow \infty} \frac{r^4}{\sqrt{-\gamma}}
\frac{\delta S_{cl}}{\delta \tilde A_\mu} = \lim_{r \rightarrow
\infty} \frac{r^4 N}{g^2} (F^{r \mu}+\frac{4\kappa_{cs} }{3}\epsilon
^{r\mu \rho \sigma \tau }A_{\rho }F_{\sigma \tau })~.
\end{equation}
Although the zeroth-order particle number current is the same as
Eq.~(\ref{zero current}), the first-order charge current
becomes~\cite{Erdmenger:2008rm,Banerjee:2008th,Son:2009tf,Hu:2011ze,Hu:2011qa}
\begin{eqnarray}
J_{(1)}^{\mu }=-\kappa P^{\mu \nu }\partial _{\nu}(\frac{\mu }{T}%
)+\sigma _{E}E^{\mu }+\sigma _{B}B^{\mu }+\xi \omega ^{\mu }+\ell
\epsilon ^{\mu \nu \rho \sigma }F_{\rho \sigma }^{\text
ext}A_{\nu}^{\text ext}\ , \label{IIIfirst order current}
\end{eqnarray}
where
\begin{eqnarray} \label{IIIfirstcoeff}
\kappa &=&  \frac{\pi ^2 T^3 r_+^7}{4 g^2M^2},~\sigma_{E} =\frac{\pi
^2 T^2 r_+^7}{4g^2 M^2},
~\sigma _{B}=-\frac{\sqrt{3} \kappa_{cs} Q (3r_{+}^4+2M)}{gMr_{+}^2},~\xi=\frac{6 \kappa_{cs} Q^2}{M},\nonumber\\
\ell&=&\frac{4 \kappa_{cs}}{3 g^2},~E^{\mu}=u^{\lambda } F ^{\rm
ext}{}_{\lambda }{}^{\mu }, ~B^{\mu }=\frac{1}{2}\epsilon ^{\mu \nu
\rho \sigma}u_{\nu}F_{\rho \sigma}^{ext},~\omega ^{\mu }=\epsilon
^{\mu \nu \rho \sigma }u_{\nu}\partial _{\rho}u_{\sigma}\ .
\end{eqnarray}
Obviously, the Chern-Simons term changes the charge current of dual
fluid, as can be seen from the last three terms in
Eq.~(\ref{IIIfirst order current}), whose coefficients are
proportional to $\kappa_{cs}$. In addition, it should be emphasized
that the external field $A_{\mu}^{ext}$ and its boundary value are
also important, as there are three terms related to the external
field $A_{\mu}^{\text ext}$, and if one chooses $A_{\mu}^{ext}(0)=0$
in Eq.~(\ref{IIExpandA}), the first-order charge current becomes
\begin{eqnarray}
J_{(1)}^{\mu }=-\kappa P^{\mu \nu }\partial _{\nu}(\frac{\mu }{T}%
)+\sigma _{E}E^{\mu }+\sigma _{B}B^{\mu }+\xi \omega ^{\mu },
\label{IIIfirst order current1}
\end{eqnarray}
which is the well-known result related to triangle
anomalies~\cite{Son:2009tf}. Moreover, if one chooses
$A_{\mu}^{ext}(0)=(C,0,0,0)$, although the first order current
Eq.~(\ref{IIIfirst order current}) becomes the same as
Eq.~(\ref{IIIfirst order current1}), the chiral magnetic conductivity is
given as $\sigma _{B}=-\frac{\sqrt{3} \kappa_{cs} Q
(3r_{+}^4+2M)}{gMr_{+}^2}+2 \ell C$, where the second term is the
extra contribution from the last term in Eq.~(\ref{IIIfirst order
current})~\cite{Hu:2011qa,Amado:2011zx}.


\section{Conclusion and discussion}
In this paper, we illustrated the algorithm of gravity/fluid
correspondence to extracting transport coefficients of dual fluid
from the bulk gravity with a simple pure gravity configuration, and
discussed its application to various bulk gravity configurations in
the presence of a $U(1)$ gauge field. On the fluid side, the
transport coefficients can be obtained by perturbing the system away
from equilibrium. On the gravity side, this can be achieved by
promoting the parameters in the boosted black brane solution to
functions of boundary coordinates. Extra correction terms are then
needed to render the boundary-coordinate-dependent metric a solution
of Einstein equations. This determines the extra correction terms
and thus the perturbative solution of the bulk gravity, from which
the transport coefficients of dual fluid can be determined using the
dictionary of gravity/fluid correspondence.

As expected, adding a $U(1)$ gauge field to the bulk gravity induces
charge current for the dual fluid. One also finds that the external
field $A_{\mu}^{ext}$ and $U(1)$ Chern-Simons term in the bulk could
affect the charge current. Some recent work investigated the dual
fluid on a finite cutoff surface and showed that the electric and
magnetic conductivity could still show up even without the external
field $A_{\mu}^{ext}$. A simple interpretation is that the
projection of $A_\mu$ onto the cutoff surface naturally introduces
its boundary value $\tilde A_\mu$ as the external field
$A_{\mu}^{ext}$~\cite{Bai:2012ci}. In addition, it should be
emphasized that the introduction of Chern-Simons term in the bulk
usually can lead to extra structures which generate anomalous
transport coefficients~\cite{Bardeen:1984pm}. For example, after
adding the Chern-Simon term of the Maxwell field in the bulk, the
conserved current $J^{\mu}$ contains additional terms related to the
anomalous magnetic and vortical
effects~\cite{Erdmenger:2008rm,Banerjee:2008th,Son:2009tf,Hu:2011ze,Hu:2011qa}.
Moreover, it has been shown that if the gravitational Chern-Simons
term is added in the bulk, a new term related to the Hall viscosity
will appear in the stress tensor of the dual
fluid~\cite{Saremi:2011ab,Chen:2011fs,Zou:2013fua}. Therefore, it
will be interesting to further investigate the effects of
Chern-Simon terms in other modified gravity configurations.


\section{Acknowledgements}

This work is supported by National Natural Science Foundation of
China (NSFC) under grant No.11105004 and Shanghai Key Laboratory of
Particle Physics and Cosmology under grant No.11DZ2230700, and
partially by grants from NSFC (No. 10821504, No. 10975168 and No.
11035008) and the Ministry of Science and Technology of China under
Grant No. 2010CB833004.


\end{document}